\documentclass[]{spie}  
\usepackage[]{graphicx,amssymb}
\def\comment#1{}\def\labell#1{\label{#1}}
\title{The quantum speed limit} \author{Vittorio
  Giovannetti\supit{a}, Seth Lloyd\supit{a,b}, and Lorenzo
  Maccone\supit{a} \skiplinehalf \supit{a}Research Laboratory of
  Electronics\\\supit{b}Department of Mechanical
  Engineering\\Massachusetts Institute of Technology\\
   77 Massachusetts Ave.,
  Cambridge, MA 02139, USA}

\begin{document}
\maketitle

\begin{abstract}
  How fast can a quantum system evolve? We answer this question
  focusing on the role of entanglement and interactions among
  subsystems. In particular, we analyze how the order of the
  interactions shapes the dynamics.
\end{abstract}

\section{Introduction}\labell{s:intro}The Hamiltonian $H$ is the
generator of the dynamics of physical systems.  In this respect, the
energy content of a system will determine its evolution time scales.
For instance, the time-energy uncertainty
relation{\cite{uncer,uncer1,mandelstam}} states that the time it takes
for a system to evolve to an orthogonal configuration in the Hilbert
space is limited by the inverse of the energy spread of the system.
Analogously, the Margolus-Levitin theorem{\cite{margolus}} relates the
same quantity to the mean energy. More generally, one can consider the
minimum time required for a system to ``rotate'' from its initial
configuration by a predetermined amount. In Ref.~{\citenum{qlimits}}
we have shown that also this time is limited by the energy and energy
spread.  More specifically, given a value of $\epsilon\in[0,1]$,
consider the time $t$ it takes for a system in the state
$|\Psi\rangle$ to evolve to a state $|\Psi(t)\rangle$ such that
\begin{eqnarray}
P(t)\equiv|\langle\Psi|\Psi(t)\rangle|^2=\epsilon
\;\labell{pidit}.
\end{eqnarray}
One can show that $t$ is always greater than the {\it quantum speed
  limit time},     \begin{eqnarray} {\cal T}_\epsilon(E,\Delta
E)\equiv\max\left(\alpha(\epsilon)\frac{\pi\hbar}{2(E-E_0)}\;,\ 
\beta(\epsilon)\frac{\pi\hbar}{2\Delta E}\right) \;\labell{qsleps},
\end{eqnarray}
where $E_0$ is the ground state energy of the system,
$E=\langle\Psi|H|\Psi\rangle$, $\Delta
E=\sqrt{\langle\Psi|(H-E)^2|\Psi\rangle}$, and $\alpha(\epsilon)$ and
$\beta(\epsilon)$ are the decreasing functions plotted in
Fig.~\ref{f:alphabeta}. If
$\epsilon=0$, from Eq.~(\ref{qsleps}) one reobtains the time-energy
uncertainty relation and the Margolus-Levitin theorem which, united,
give the following lower bound for the time it takes for a system to
evolve to an orthogonal configuration, \begin{eqnarray}
  t\geqslant{\cal T}_0(E,\Delta
  E)\equiv\max\left(\frac{\pi\hbar}{2(E-E_0)}\;,\ 
    \frac{\pi\hbar}{2\Delta E}\right) \;\labell{qsl}.
\end{eqnarray} In the rest of the paper we will focus on this case
only as a followup of a previous paper of ours{\cite{role}}. It is
worth mentioning that when the system is a mixed state $\varrho$, the
quantum speed limit time ${\cal T}_\epsilon$ is the minimum time it
takes for the system to evolve to a configuration $\varrho(t)$ such
that the fidelity{\cite{jozsa}}
\begin{eqnarray} 
F(\varrho,\varrho(t))
\equiv\left\{\mbox{Tr}\left[\sqrt{\sqrt{\varrho}{\varrho(t)}\sqrt{\varrho}}\right]\right\}^2
\;\labell{fid}
\end{eqnarray}
between these two states is equal to $\epsilon$ (see
Ref.~{\citenum{qlimits}} for details).  

In addition to giving upper bounds to the evolution ``speed'' of a
system, the quantity ${\cal T}_\epsilon$ is useful to analyze the role
of the entanglement among subsystems in the dynamics of composite
systems: in Sec.~\ref{s:ent} we review some of the material presented
in Ref.~\citenum{role} pointing out the differences between entangled
and separable pure states in achieving the quantum speed limit bound
in the context of non-interacting subsystems; in Sec.~\ref{s:int}, on
the other hand, we address the case of interacting subsystems and
analyze in detail how the order of interaction influences the
dynamical ``speed'' of composite systems.

\begin{figure}[hbt]
\begin{center}
   \begin{tabular}{c}
   \includegraphics[height=7cm]{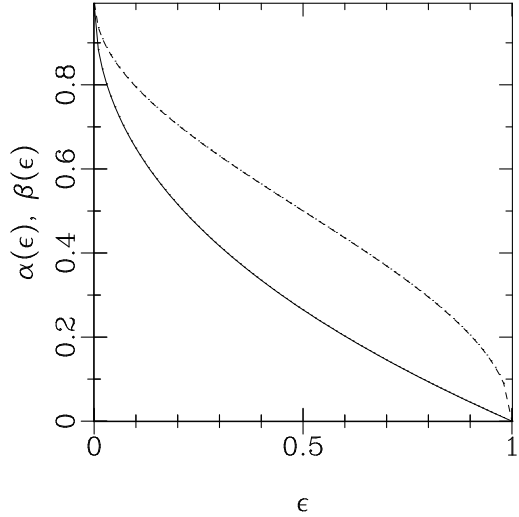}
   \end{tabular}
   \end{center}
   \caption[example] 
   { \labell{f:alphabeta} The functions $\alpha(\epsilon)$ (continuous
     line) and $\beta(\epsilon)$ (dashed line) of Eq.~(\ref{qsl}). The
     analytical expression of $\beta$ is
     $\beta(\epsilon)=2\arccos(\sqrt{\epsilon})/\pi$; the explicit
     analytical expression for $\alpha(\epsilon)$ is not
     known{\cite{qlimits}}, even though one can evaluate this function
     numerically showing that
     $\alpha(\epsilon)\simeq\beta^2(\epsilon)$.}
   \end{figure}

\section{Entanglement and dynamics: non-interacting subsystems}\labell{s:ent}
We show that pure separable states cannot reach the quantum speed
limit unless all energy resources are devoted to one of the
subsystems. This is no longer true if the initial state is entangled:
a cooperative behavior is induced such that the single subsystems
cannot be regarded as independent entities.

Consider a system composed of $M$ parts which do not interact.  The
Hamiltonian for such a system is
\begin{eqnarray} H=\sum_{i=1}^MH_i \;\labell{hamil},
\end{eqnarray}
where $H_i$ is the free Hamiltonian of the $i$th subsystem.  Without
loss of generality, we assume that each $H_i$ term (and hence $H$) has
zero ground state energy. A separable pure state has the form
\begin{eqnarray}
|\Psi_{sep}\rangle=|\psi_1\rangle_1\otimes\cdots\otimes|\psi_M\rangle_M
\;\labell{stcl},
\end{eqnarray}
where $|\psi_i\rangle_i$ is the state of the $i$-th subsystem which
has energy ${\cal E}_i$ and spread $\Delta{\cal E}_i$.  Since there is
no interaction, the vector $|\Psi_{sep}\rangle$ remains factorizable
at all times, i.e. \begin{eqnarray}
  |\Psi_{sep}(t)\rangle=|\psi_1(t)\rangle_1\otimes\cdots\otimes|\psi_M(t)\rangle_M
\;\labell{ppsi}.
\end{eqnarray} This state becomes orthogonal to its initial
configuration (\ref{stcl}) if at least one of the subsystems
$|\psi_i\rangle_i$ evolves to an orthogonal state. The time ${\cal
  T}_\perp$ employed by this process is limited by the energies and
the energy spreads of each subsystem, through Eq.~(\ref{qsl}). By
considering the time corresponding to the ``fastest'' subsystem, the
quantity ${\cal T}_\perp$ has to satisfy the inequality
\begin{eqnarray} {\cal
T}_\perp\geqslant\max\left(\frac{\pi\hbar}{2{\cal E}_{max}}\;,\
\frac{\pi\hbar}{2\Delta {\cal E}_{max}}\right)
\;\labell{qsl1},
\end{eqnarray}
where ${\cal E}_{max}$ and $\Delta{\cal E}_{max}$ are the maximum
values of the energy and energy spreads of the $M$ subsystems.  For
the state $|\Psi_{sep}\rangle$, the total energy and the total energy spread
are \begin{eqnarray} E&=&\sum_i{\cal E}_i\geqslant {\cal
    E}_{max},\\\Delta E&=&\left[{\sum_i\Delta{\cal
E}^2_i}\right]^{\frac 12}\geqslant\Delta{\cal E}_{max}
\;\labell{en}.
\end{eqnarray} This implies that the bound imposed by Eq.~(\ref{qsl1}) is
always greater or equal than ${\cal T}_0(E,\Delta E)$ of
Eq.~(\ref{qsl}), attaining equality only when ${\cal E}_{max}=E$ or
$\Delta{\cal E}_{max}=\Delta E$, e.g. when one of the subsystems has
all the energy or all the energy spread of the whole system. In both
these cases, only one such subsystem evolves in time and the remaining
$M-1$ are stationary. The gap between the bound (\ref{qsl1}) for
separable pure states and the bound (\ref{qsl}) for arbitrary states
reaches its maximum value for systems that are {\it homogeneous} in
the energy distribution, i.e. such that ${\cal E}_{max}=E/M$ and
$\Delta{\cal E}_{max}=\Delta E/\sqrt{M}$. In this case,
Eq.~(\ref{qsl1}) implies that, for factorizable states, ${\cal
  T}_\perp/{\cal T}_0(E,\Delta E)\geqslant\sqrt{M}$.

\subsection{Entangled subsystems.}\labell{s:sub}
In order to show that the bound {\it is} indeed achievable when
entanglement is present, consider the following entangled state
\begin{eqnarray} |\Psi_{ent}\rangle=\frac
1{\sqrt{N}}\sum_{n=0}^{N-1}|n\rangle_1\otimes\cdots\otimes|n\rangle_M
\;\labell{stent},
\end{eqnarray}
where $|n\rangle_i$ is the energy eigenstate (of energy
$n\hbar\omega_{0}$) of the $i$-th subsystem and $N\geqslant 2$. The
state $|\Psi_{ent}\rangle$ is homogeneous since each of the $M$
subsystems has energy and spread \begin{eqnarray}
  {\cal E}&=&\hbar\omega_{0} (N-1)/2\;,\\
  \Delta {\cal
    E}&=&\hbar\omega_{0}\sqrt{N^2-1}/(2\sqrt{3})\;\labell{enn1}.
\end{eqnarray}
On the other hand, the total energy and energy spread are given by
$E=M{\cal E}$ and $\Delta E=M\Delta {\cal E}$ respectively and the
quantum speed limit time (\ref{qsl1}) becomes,
\begin{eqnarray}
{\cal T}_0=\frac{\sqrt{3}\;\pi}{M\sqrt{N^2-1}\; {\omega_0}}
\;\labell{qsl1ent}.
\end{eqnarray}
   The scalar product of $|\Psi_{ent}\rangle$ with its
time evolved $|\Psi_{ent}(t)\rangle$ gives
\begin{eqnarray} P(t)=\left|\frac
1N\sum_{n=0}^{N-1}e^{-inM\omega_{0} t}\right|^2
\;\labell{evoluz},
\end{eqnarray}
where the factor $M$ in the exponential is a peculiar signature of the
energy entanglement.  The value of ${\cal T}_\perp$ for the state
$|\Psi_{ent}\rangle$ is given by the smallest time $t\geqslant 0 $ for
which $P(t)=0$, i.e. ${\cal T}_\perp=2\pi/(NM\omega_{0})$.  This
quantity is smaller by a factor $\sim\sqrt{M}$ than what it would be
for homogeneous separable pure states with the same value of $E$ and
$\Delta E$, as can be checked through Eq.~(\ref{qsl1}). Moreover,
${\cal T}_\perp/{\cal T}_0(E,\Delta E)$ does not depend on the number
$M$ of subsystems and, for any value of the number of energy levels
$N$, is always close to one (see Fig.~\ref{f:rapporto}). In
particular, for $N=2$, the system achieves the quantum speed limit,
i.e. ${\cal T}_\perp={\cal T}_0(E,\Delta E)$. 

\begin{figure}[hbt]
\begin{center}
   \begin{tabular}{c}
   \includegraphics[height=7cm]{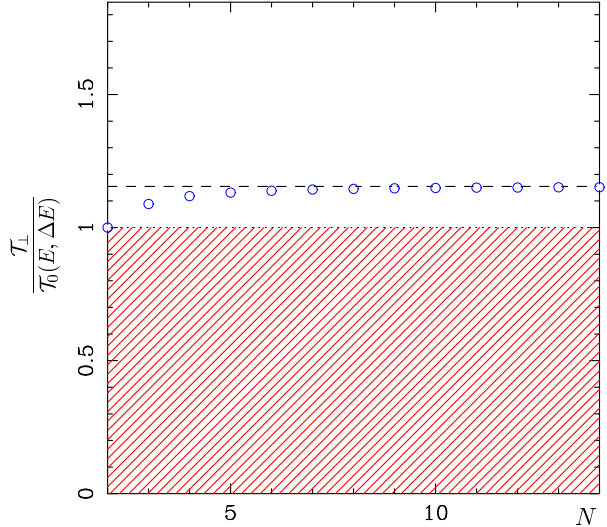}
   \end{tabular}
   \end{center}
   \caption[example] 
   { \labell{f:rapporto} Plot of the ratio between the minimum time
     ${\cal T}_\perp$ it takes for system to evolve to an orthogonal
     configuration starting from the state $|\Psi_{ent}\rangle$ of
     Eq.~(\ref{stent}) and the quantum speed limit time ${\cal
       T}_0(E,\Delta E)$ of Eq.~(\ref{qsl1ent}) as a function of the
     number $N$ of eigenstates of each subsystem. This ratio does not
     depend on the number of subsystems $M$ and for all $N$ is
     upper-bounded by its asymptomatic value $2/\sqrt{3}$ (dashed
     line). }
\end{figure}

The above discussion shows that, for non interacting subsystems, pure
separable states can reach the quantum speed limit only in the case of
highly asymmetric configurations where one of the subsystems evolves
to an orthogonal configuration at the maximum speed allowed by its
energetic resources, while the other subsystems do not evolve. In the
other cases (for instance the homogeneous configuration discussed
above) entanglement is a necessary condition to achieve the bound.
This, of course, does not imply that all entangled states evolve
faster than their unentangled counterparts. In
Refs.~{\citenum{role,qlimits}} these arguments have been extended, with
some caveats, to the case of mixed states.

\section{Interactions and entanglement}\labell{s:int}
In the case in which there are interactions among different
subsystems, the Hamiltonian is \begin{eqnarray}
H=\sum_{i=1}^MH_i+H_{int}
\;\labell{intham},
\end{eqnarray}
where $H_{int}$ acts on the Hilbert spaces of more than one subsystem.
Two situations are possible: either $H_{\rm int}$ does not introduce
any entanglement in the initial state of the system or $H_{\rm int}$
builds up entanglement among subsystems. In the first case, no
correlations among the subsystems are created so that each subsystem
evolves independently as in Eq.~(\ref{ppsi}), unless entanglement was
present initially. Since this type of evolution can always be
described as determined by an interaction-free effective Hamiltonian,
the results of the previous section apply: i.e. the system reaches the
quantum speed limit only if entanglement is present initially or if
all energy resources are devoted to a single subsystem.  In the second
case, when $H_{\rm int}$ builds up entanglement, the system may reach
the bound even though no entanglement was already present initially.
In fact, one can tailor suitable entangling Hamiltonians that speed up
the dynamical evolution even for initial homogeneous separable pure
states.  In order to reach the bound, however, the interaction must
{\it i)} connect all the qubits and {\it ii)} be sufficiently
strong{\cite{role}}.  A recent proposal of one of us{\cite{sethch}}
uses this effect to increase the communication rate of a qubit channel
by a factor $\sqrt{M}$ over a communication channel composed of $M$
independent parallel channels which uses the same resources.

The order $K$ of the interaction (i.e. the number of subsystems that
are involved in a single vertex of interaction) also plays an
important role. The case analyzed in Ref.~\citenum{role} assumes an
$M$th order interaction to show that a separable pure state can be
brought to the quantum speed limit even when energy resources are
equally distributed among subsystems. Here we want to discuss in some
detail what happens for $K<M$. For any given $K$, a rich variety of
configurations are possible depending on how the interaction is
capable of constructing entanglement between subsystems. The simplest
example is an Ising-like model where there is a $K$-th order coupling
between neighbors in a chain of qubits. In this context a $\sqrt{K}$
improvement over the non-interacting case is achievable. We now work
out this model in detail. Consider a system of $M$ qubits governed by
the following Hamiltonians:
\begin{eqnarray}
  H_i&=&\hbar\omega_0(1-\sigma_x^{(i)})\qquad (i=1,\cdots,M)\;,\labell{interaz}\\
  H_{int}&=&\hbar\omega\sum_{j=1}^{Q}(1-S_j) \;\labell{interaz1},
\end{eqnarray}
where $\sigma_x^{(i)}$ is the Pauli operator
$|0\rangle\langle 1|+|1\rangle\langle 0|$ acting on the $i$th spin,
$Q$ is the number of interaction terms and \begin{eqnarray}
  S_j\equiv\sigma_x^{(i_{1j})}\otimes\sigma_x^{(i_{2j})}\otimes\cdots\otimes\sigma_x^{(i_{Kj})}
  \qquad(j=1,\cdots,Q) \;\labell{sigmajay},
\end{eqnarray}
with $i_{lj}=1,\cdots,M$ for $l=1,\cdots,K$ and $i_{lj}\neq i_{l'j}$
for $l\neq l'$.  The operator $S_j$ is a $K$th order interaction
Hamiltonian that connects the spins $i_{1j}$ through $i_{Kj}$. The
free Hamiltonians $H_i$ rotate each spin along the $x$ axis at a
frequency $\omega_0$, while the interaction $H_{int}$ collectively
rotates the $Q$ blocks of $K$ spins at a frequency $\omega$. For ease
of calculation, we assume that each spin belongs only to two of the
$Q$ interacting groups and that no spin is left out. This enforces a
symmetry condition on the qubits (i.e. all qubits are subject to the
same interactions) and implies that $S_1S_2\cdots S_Q=1$.  Consider an
initial state in which all qubits are in $\sigma_z=|0\rangle\langle
0|-|1\rangle\langle 1|$ eigenstates, such as
$|\Psi\rangle=|0\rangle_1\otimes\cdots\otimes|0\rangle_M$.  This state
is clearly homogeneous and has energy characteristics
\begin{eqnarray}
  E&=&\hbar\omega_0\;M+\hbar\omega\;Q\\
  \Delta E&=&\sqrt{(\hbar\omega_0)^2M+(\hbar\omega)^2Q}
  \;\labell{enn},
\end{eqnarray}
so that, in the strong interaction regime{\footnote{In
    Ref.~\citenum{role} we have shown that the strong interaction
    regime is the only one that allows to reach the quantum speed
    limit, since only in this regime one can build strong enough
    correlations among all subsystems.}} $\ \omega\gg\omega_0$, its
quantum speed limit time (\ref{qsl}) is{\footnote{Notice that the
    ground state energy $E_0$ associated with the Hamiltonians
    (\ref{interaz}) and (\ref{interaz1}) is null.}} \begin{eqnarray}
  {\cal T}_0(E,\Delta E)\simeq\frac \pi{2\omega\sqrt{Q}}
  \;\labell{qslint}.
\end{eqnarray}
Since $[H_i,H_{int}]=0$ for all $i$ and $[S_j,S_j']=0$ for all $j,j'$,
the time-evolved state has the form
\begin{eqnarray}
|\Psi(t)\rangle=U_0(t)U_{int}(t)|\Psi\rangle\;\labell{biancaneve},
\end{eqnarray}with
\begin{eqnarray}
  U_0(t)=e^{-i\omega_0t}\bigotimes_{i=1}^M(\cos\omega_0t+i\;
  \sigma_x^{(i)}\sin\omega_0t)\;,\qquad
  U_{int}(t)=e^{-i\omega t} \prod_{j=1}^Q(\cos\omega
  t+i\;S_j\sin\omega t)\;.  \;\labell{evolt}
\end{eqnarray}
For $\omega\gg\omega_0$, this yields \begin{eqnarray}
  P(t)=|\langle\Psi|\Psi(t)\rangle|^2\simeq|(\cos\omega
  t)^Q+(i\;\sin\omega t)^Q|^2 \;\labell{cos}.
\end{eqnarray}
If $Q$ is an odd number or twice an even number, $P(t)$ is never zero:
in this case the interaction does not help in reaching the quantum
speed limit{\footnote{This is an artifact of our model. One can devise
    different configurations (e.g. dropping the qubits symmetry
    requirement) where these limitations do not apply.}}.  However, if
$Q$ is twice an odd number, then $P(t)=0$ has solution and the minimum
time $t$ for which this condition is satisfied is ${\cal
  T}_\perp=\pi/(4\omega)$.  The ratio between this quantity and the
quantum speed limit time (\ref{qslint}) is, then, ${\cal
  T}_\perp/{\cal T}_0(E,\Delta E)=\sqrt{Q}/2$. We now need to connect
the quantity $Q$ to the order $K$ of the interaction and the number
$M$ of spins. A simple way to visualize the interactions among qubits
is to arrange them into polygonal structures, as in
Fig.~\ref{f:cubetti}, where each line represents one of the $Q$
interactions $S_j$. Using this representation, it is easy to verify
that $Q=L+L(K-2)/K$, where $L=M/(K-1)$ is the number of sides of the
polygon. This means that
\begin{eqnarray} \frac{{\cal T}_\perp}{{\cal T}_0(E,\Delta
    E)}=\sqrt{\frac{M}{2K}}\;\labell{rapp},
\end{eqnarray}
which gives a $\sqrt{K}$ improvement over the case of non-interacting
subsystems for homogeneous separable configurations.

\begin{figure}[hbt]
\begin{center}
   \begin{tabular}{c}
   \includegraphics[height=7cm]{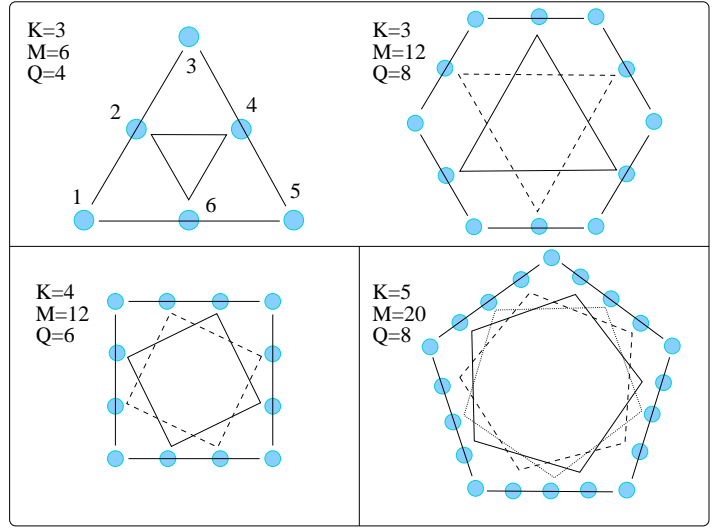}
   \end{tabular}
   \end{center}
   \caption[example] 
   { \labell{f:cubetti} Relation between the order $K$ of the
     interaction, the number $M$ of qubits (i.e. the gray dots) and
     the number $Q$ of interaction terms for the example discussed in
     the text. The qubits can be organized in regular polygons with
     $L$ sides, each containing $K-1$ qubits: here we show some of the
     possible configurations. The lines (continuous, dashed or dotted)
     represent the $Q$ interactions $S_j$ ($j=1,\cdots,Q$): each qubit
     is involved in two interaction terms. For example, in the first
     structure (with six qubits), the qubit number 2 interacts with
     qubits 1,3 and with 4,6. Among the examples plotted here, only the
     case $K=4,\;M=12,\;Q=6$ satisfies Eq.~(\ref{rapp}).}
\end{figure} 

\section{Conclusions}\labell{s:concl}
In summary, we have assembled all the dynamical limitations connected
to the energy of a system in a {\it quantum speed limit} relation,
Eq.~(\ref{qsleps}), that establishes how fast any physical system can
evolve given its energy characteristics.  We have proved that
homogeneous separable pure states cannot exhibit speedup, while at
least one entangled case that exhibits speedup exists. This suggests a
fundamental role of energy entanglement in the dynamical evolution of
composite systems. Moreover, we analyzed the role of interactions with
symmetric configurations in which each subsystem interacts directly
with $K-1$ other subsystems. In this case, we have shown that a
$\sqrt{K}$ enhancement is possible over the non-interacting
unentangled case with energy shared among subsystems.

\acknowledgments
This work was funded by the ARDA, NRO, NSF, and by ARO under a MURI
program.

\bibliography{report}   
\bibliographystyle{spiebib}   
\end{document}